\documentclass[a4paper,fleqn,usenatbib]{mnras}

\usepackage{newtxtext,newtxmath}

\usepackage[T1]{fontenc}
\usepackage{ae,aecompl}

\usepackage{graphicx}	
\usepackage{amsmath}	
\usepackage{amssymb}	
\usepackage[english]{babel}
\usepackage{blindtext}
\usepackage{verbatim}

\newcommand{\DLB}{\textsc{DLB07}}
\newcommand{\LGal}{\textsc{Lgalaxies}}
\newcommand{\morgana}{\textsc{Morgana}}
\newcommand{\sage}{\textsc{Sage}}
\newcommand{\ysam}{\textsc{ySAM}}
\newcommand{\galics}{\textsc{GalICS-2.0}}
\newcommand{\sag}{\textsc{Sag}}
\newcommand{\mice}{\textsc{Mice}}
\newcommand{\galform}{\textsc{Galform}}
\newcommand{\rockstar}{\textsc{Rockstar}}
\newcommand{\eagle}{\textsc{Eagle}}
\newcommand{\Fig}[1]{Figure~\ref{#1}}
\newcommand{\hMpc}{{\ifmmode{h^{-1}{\rm Mpc}}\else{$h^{-1}$Mpc}\fi}}
\newcommand{\hMsun}{{\ifmmode{h^{-1}{\rm
{M_{\odot}}}}\else{$h^{-1}{\rm{M_{\odot} }}$}\fi}}

\title[Cosmic CARNage II]{Cosmic CARNage II: the evolution of
the galaxy stellar mass function in observations and galaxy formation models}

\author[R. Asquith et al.]
      {Rachel Asquith,$^{1}$\thanks{E-mail: rachel.asquith@nottingham.ac.uk}
      Frazer R. Pearce,$^{1}$				
      Omar Almaini,$^{1}$				
      Alexander Knebe,$^{2,3}$				
      \newauthor
      Violeta Gonzalez-Perez,$^{4,5}$			
      Andrew Benson,$^{6}$    				
      Jeremy Blaizot,$^{7,8,9}$ 			
      Jorge Carretero,$^{10,11}$ 			
      \newauthor
      Francisco J. Castander,$^{10}$	 		
      Andrea Cattaneo,$^{12}$				
      Sof\'ia A. Cora,$^{13,14}$				
      Darren J. Croton,$^{15}$			        
      \newauthor
      Julien E. Devriendt,$^{16}$			
      Fabio Fontanot,$^{17}$				
      Ignacio D. Gargiulo,$^{13,14}$	 		
      Will Hartley,$^{18}$				
      \newauthor
      Bruno Henriques,$^{19}$		        	
      Jaehyun Lee,$^{20}$				
      Gary A. Mamon,$^{21}$				
      Julian Onions,$^{1}$				
      \newauthor
      Nelson D. Padilla,$^{22,23}$ 			
      Chris Power,$^{24}$				
      Chaichalit Srisawat,$^{25}$ 	                
      Adam R. H. Stevens,$^{15,24}$			
      \newauthor
      Peter A. Thomas,$^{25}$			 	
      Cristian A. Vega-Mart\'inez,$^{13}$ 		
      Sukyoung K. Yi$^{26}$			        
\\
\\
$^{1}$School of Physics \& Astronomy, University of Nottingham, Nottingham NG7
2RD, UK\\
$^{2}$Departamento de F\'isica Te\'{o}rica, M\'{o}dulo 15, Facultad de Ciencias,
Universidad Aut\'{o}noma de Madrid, 28049 Madrid, Spain\\
$^{3}$Centro de Investigaci\'{o}n Avanzada en F\'isica Fundamental (CIAFF),
Facultad de Ciencias, Universidad Aut\'{o}noma de Madrid, 28049 Madrid, Spain\\
$^{4}$Institute for Computational Cosmology, Department of Physics, University
of Durham, South Road, Durham, DH1 3LE, UK\\
$^{5}$Institute of Cosmology \& Gravitation, University of Portsmouth, Dennis
Sciama Building, Portsmouth, PO1 3FX, UK\\
$^{6}$Carnegie Observatories, 813 Santa Barbara Street, Pasadena, CA 91101,
USA\\
$^{7}$Universit\`e de Lyon, Lyon, F-69003, France\\
$^{8}$Universit\`e Lyon 1, Observatoire de Lyon, 9 avenue Charles Andr\`e,
Saint-Genis Laval, F-69230, France\\
$^{9}$CNRS, UMR 5574, Centre de Recherche Astrophysique de Lyon ; Ecole Normale
Sup\`erieure de Lyon, Lyon, F-69007, France\\
$^{10}$Institut de Ci\`encies de l'Espai, IEEC-CSIC, Campus UAB, 08193
Bellaterra, Barcelona, Spain\\
$^{11}$Port d{'}Informaci\'{o} Cient\'{i}fica (PIC) Edifici D, Universitat
Aut\`{o}noma de Barcelona (UAB), E-08193 Bellaterra (Barcelona), Spain.\\
$^{12}$GEPI, Observatoire de Paris, CNRS, 61, Avenue de l'Observatoire 75014,
Paris  France\\
$^{13}$Instituto de Astrof\'isica de La Plata (CCT La Plata, CONICET, UNLP),
Paseo del Bosque s/n, B1900FWA, La Plata, Argentina.\\
$^{14}$Facultad de Ciencias Astron\'omicas y Geof\'{\i}sicas, Universidad
Nacional de La Plata, Paseo del Bosque s/n, B1900FWA, La Plata, Argentina\\
$^{15}$Centre for Astrophysics and Supercomputing, Swinburne University of
Technology, Hawthorn, Victoria 3122, Australia\\
$^{16}$Astrophysics, University of Oxford, Denys Wilkinson Building, Keble Road,
Oxford, OX1\,3RH, UK\\
$^{17}$INAF - Astronomical Observatory of Trieste, via Tiepolo 11, I-34143
Trieste, Italy\\
$^{18}$Department of Physics and Astronomy, University College London, Gower
Street, London WC1E 6BT\\
$^{19}$Max-Planck-Institut f\"ur Astrophysik, Karl-Schwarzschild-Str. 1, 85741
Garching b. M\"unchen, Germany\\
$^{20}$Korea Institute for Advanced Study, 85 Hoegiro Dongdaemun-gu, Seoul 02455
Korea\\
$^{21}$Institut d'Astrophysique de Paris (UMR 7095: CNRS \& UPMC), 98 bis Bd
Arago, F-75014 Paris, France\\
$^{22}$Instituto de Astrofisica, Universidad Catolica de Chile, Santiago,
Chile\\
$^{23}$Centro de Astro-Ingenieria, Universidad Catolica de Chile, Santiago,
Chile\\
$^{24}$International Centre for Radio Astronomy Research, University of Western
Australia, 35 Stirling Highway, Crawley, \\ Western Australia 6009, Australia\\
$^{25}$Department of Physics \& Astronomy, University of Sussex, Brighton, BN1
9QH, UK\\
$^{26}$Department of Astronomy and Yonsei University Observatory, Yonsei
University, 03722 Seoul, Republic of Korea\\
}

\date{Accepted 2018 July 09. Received 2018 July 09; in original form 2017
December 19}

\pubyear{2018}

\begin{document}
\label{firstpage}
\pagerange{\pageref{firstpage}--\pageref{lastpage}}
\maketitle
\clearpage

\begin{abstract}
We present a comparison of the observed evolving galaxy stellar mass functions
with the predictions of eight semi-analytic models and one halo occupation
distribution model. While most models are able to fit the data at low redshift,
some of them struggle to simultaneously fit observations at high redshift. We
separate the galaxies into `passive' and `star-forming' classes and find that
several of the models produce too many low-mass star-forming galaxies at high
redshift compared to observations, in some cases by nearly a factor of 10 in the
redshift range $2.5 < z < 3.0$. We also find important differences in the
implied mass of the dark matter haloes the galaxies inhabit, by comparing with
halo masses inferred from observations. Galaxies at high redshift in the models
are in lower mass haloes than suggested by observations, and the star formation
efficiency in low-mass haloes is higher than observed. We conclude that many of
the models require a physical prescription that acts to dissociate the growth of
low-mass galaxies from the growth of their dark matter haloes at high redshift.
\end{abstract}

\begin{keywords}
methods:numerical -- galaxies:haloes -- galaxies: evolution -- cosmology:theory
-- dark matter
\end{keywords}



\section{Introduction} \label{sect:intro}
Locally, low-mass galaxies tend to be disky, blue and star-forming, whereas
high-mass galaxies are more likely to be spheroidal, red and passive
\citep[e.g.][]{ref:Kennicutt98, ref:Strateva01, ref:Kauffmann03, ref:Baldry04}.
At high redshift $(z>1)$ we also observe this bimodality in the galaxy
population \citep{ref:Kovac14, ref:Cirasuolo07}, but do not definitively know
the mechanisms by which these galaxies evolve into the populations we observe
locally. Various mechanisms have been suggested to move galaxies from the `blue
cloud' to the `red sequence' and shut off their star formation in a process
known as `quenching'. Potential quenching mechanisms include environmental
effects and feedback from active galactic nuclei (AGN) at high masses \citep[for
a review see][]{ref:Benson10}, but these processes are still not fully
understood.

One way to study this problem is to directly observe galaxies forming and
evolving in the distant Universe. At high redshift $(z>1)$, deep near-infrared
observations are vital to select galaxies by rest-frame optical light. Selecting
high-redshift galaxies using optical imaging will introduce strong biases
against dusty galaxies or those with evolved (i.e. passive) stellar populations
\citep[e.g.][]{ref:Cowie96}. It is only recently that deep near-infrared surveys
have been conducted with the required depth and area to produce large galaxy
samples at high redshift, sufficient to allow accurate determinations of the
galaxy stellar mass function while minimising the influence of cosmic variance. 
In particular, the UKIDSS Ultra Deep Survey (UDS) \citep[][Almaini et al. in
prep.]{ref:Lawrence07} and UltraVISTA \citep{ref:McCracken12} are now deep
enough to detect typical (i.e. $M^*$) galaxies to $z \sim 3$, over large volumes
of the distant Universe ($\sim 100 \times 100$ projected comoving Mpc at $z=3$).
Using these surveys, we can directly test model predictions for the build-up of
the galaxy populations, rather than inferring their evolution by extrapolating
back in time. However, each galaxy is only being seen at one point in its life
and we cannot infer the full evolutionary history.

In order to get a cohesive picture of what happens to galaxies throughout their
lives, one approach is to link a population of galaxies at high redshift to a
population at low redshift that could be their descendants. This can be done by
selecting galaxies at a constant comoving number density when ranked by mass or
luminosity \citep{ref:Mundy15}. This method was partly motivated by the need to
overcome `progenitor bias', where new young star-forming galaxies enter the
sample at low redshift that are not present at high redshift
\citep{ref:Shankar15}. Not accounting for this bias correctly can lead to a poor
selection of the set of galaxies being connected as progenitor and descendent
and therefore incorrect conclusions being drawn about their evolution.

A powerful method to trace galaxies through redshift is to use semi-analytic
models (SAMs) \citep[for a review see][]{ref:Benson10, ref:Somerville15}, a type
of galaxy formation model in which simple analytic prescriptions (in connection
with merger trees from either cosmological simulations or extended
Press-Schechter formalisms) are used to model the physical processes occurring
during galaxy formation and evolution. These models are able to evolve the same
population of galaxies through redshift and connect them without the limitations
of observational methods. These models are also computationally inexpensive, so
can be used to simulate large volumes and produce large catalogues of galaxies
with which to compare observational data. By comparing the models to key
observables, e.g. the evolution of the stellar mass function (SMF), we can learn
about the physics of galaxy formation. If models are not able to reproduce
observational results it may mean that they are missing key physics which is
important in galaxy formation and evolution. Model galaxies can also be
separated into `star-forming' and `passive' types, to test for the quenching
processes which transform galaxies from star-forming to passive.

While it has been shown that SAMs are able to reproduce the SMF at $z=0$, they
struggle to simultaneously match observations at both low and high redshift
\citep[e.g.][]{ref:Fontanot09,ref:Weinmann12,ref:Guo11,ref:Knebe15}. This has
only become clearer in recent years as observational surveys have been able to
probe down to lower masses as well as probing to higher redshifts. Observational
evidence appears to point towards a seemingly `anti-hierarchical' formation
scenario where high-mass galaxies form earlier with their abundance changing
little from $z \sim 1$ to the present day, whereas there is a rapid evolution in
the number of low-mass galaxies at late times \citep[e.g.][]{ref:Fontana04,
ref:Fontana06, ref:Faber07, ref:Pozzetti07, ref:Marchesini09, ref:Marchesini10,
ref:Pozzetti10, ref:Ilbert10, ref:Ilbert13, ref:Muzzin13}. This is sometimes
referred to as `mass assembly downsizing' \citep{ref:Cowie96, ref:Cimatti06,
ref:Lee13}.

After much work understanding both AGN feedback and the mass assembly of
high-mass galaxies \citep[e.g.][]{ref:Benson03, ref:DiMatteo05, ref:Bower06,
ref:Croton06}, models are now able to reproduce the high-mass end of the galaxy
stellar mass function over a range of redshifts. However, models still typically
overproduce the number of low-mass galaxies at high redshift. The main reason
for this discrepancy appears to be that galaxies in the models follow the growth
of their dark matter haloes too closely \citep{ref:Weinmann12, ref:Somerville15,
ref:Guo16}. Halo mass growth is the main driver of gas accretion rate in
galaxies, which then in turn drives the star formation rate. The star formation
history then traces the dark matter mass accretion history, which, in the
favoured $\Lambda$CDM structure formation scenario, is approximately
self-similar for haloes of different masses. However, in the real Universe it
appears that there is not such a tight correlation \citep{ref:White15,
ref:Guo16}.

This excess of low-mass galaxies at high redshift was investigated by
\citet{ref:Fontanot09}, who found that in three different SAMs, galaxies in the
mass range $9 < \mathrm{log}(M_*/\mathrm{M_\odot}) < 11$ form too early and have
little ongoing star formation at late times. They concluded that the physical
processes operating on these mass scales, such as supernova feedback, needed a
re-think. \citet{ref:Weinmann12} later used two SAMs and two cosmological
hydrodynamical simulations and examined the evolution of the observed number
density of galaxies. They found that although the models fit well at $z = 0$,
the low-mass galaxies were formed at early times. They conclude that as the
current form of feedback is mainly dependent on host halo mass and time, it is
unlikely to be able to separate the growth of galaxies from the growth of their
dark matter haloes.

Monte Carlo Markov Chain (MCMC) methods were used by \citet{ref:Henriques13} 
in an attempt to fit the stellar mass function at all redshifts, but they could
not find a single set of parameters that allowed this. They then changed the
reincorporation timescale for ejected gas to be inversely proportional to halo
mass and independent of redshift and found that they were able to fit observed
numbers of low-mass galaxies from $0 < z < 3$. However, the passive fraction of
low-mass galaxies was still too high.  Their model was later updated further in
\citet{ref:Henriques15} where they also reduce ram-pressure stripping in
low-mass haloes, make radio-mode AGN feedback more efficient at low redshift,
and reduce the gas surface density threshold for star formation. They then find
that their model reproduces the observed abundance and passive fraction of
low-mass galaxies, both at high and low redshift.

Another attempt to solve this problem was by \citet{ref:White15}, who tried
three different physically motivated methods to decouple the accretion rate in
galaxies from their star formation rate. They found that changing the gas
accretion to be less efficient in low-mass haloes at early times and increasing
the dependence of stellar feedback on halo mass at high redshift were the most
successful at qualitatively matching the evolution of the number density of
low-mass galaxies. However, they allow these functions to scale with halo mass
and redshift in an arbitrary way which may not be physically motivated.
\citet{ref:Hirschmann16} also investigated this problem with their model and
found that they improved their agreement with observations by either reducing
the gas ejection rate with cosmic time or varying the reincorporation timescale
with halo mass, classed as `ejective' and `preventative' feedback schemes
respectively. Although their results improve from their fiducial model, they
still find too many low-mass, red, old galaxies between $0.5 < z < 2.0$.

However, the effect of adjusting certain physical prescriptions can be vastly
different between models. \citet{ref:White15} investigated what effect
replicating the changes in \citet{ref:Henriques13} had on their own model, but
found that it did not make much difference to the observed number density of
low-mass galaxies. They conclude that this is due to the sensitivity of the
results to how the gas reservoirs are tracked and treated in the different
codes. \citet{ref:Croton16} also had similar problems with this approach and
found that it did not solve the problems with fitting the stellar mass
function. This presents difficulties to the modelling community, as it means
that different models may require different changes to get them to match
the observed evolution.

It is also possible to try and match the galaxy stellar mass function at all
redshifts without changing the physics involved in the model. For example,
\citet{ref:Rodrigues17} used \galform\ to identify a small region of parameter
space where the model matched the observational data out to $z=1.5$, without
needing to adapt any of the physics involved. They found that the parameters
controlling the feedback processes were most strongly constrained, suggesting
that these processes are important when fitting the evolution of the galaxy
stellar mass function.

Halo occupation distribution (HOD) models, rather than modelling the physical
processes that we think go into galaxy formation, use statistical methods to
match galaxies to their corresponding dark matter haloes
\citep[e.g.][]{ref:Berlind02, ref:Zheng05}. As these models are applied
independently at each redshift, the evolution of each galaxy is not tracked,
although they can be connected to their progenitors and descendants via dark
matter merger trees. HOD models by design are able to reproduce the SMF at each
redshift and are therefore able to reproduce the population of galaxies at any
given time. This type of model is a very useful tool for learning about the
relationship between galaxies and their host dark matter haloes and how this
changes as a function of redshift. For example, \citet{ref:Berlind03} found that
low-mass haloes are mainly populated by young galaxies and high-mass haloes by
older galaxies.

Galaxy formation models such as HODs and SAMs must be calibrated using
observational datasets. Varying the calibration dataset, {\it even for the same
model} may produce significantly different catalogues. Essentially, the
calibration datasets introduce tension, and it may not be possible for a single
model to fit all the required observational datasets simultaneously. This could
be because the model lacks some of the required physics or that the underlying
observational datasets are incomplete or are physically incompatible with each
other.

In the Cosmic CARNage mock galaxy comparison project \citep[][hereafter referred
to as Paper I]{ref:Knebe18} we sought to address some of these issues by
requiring the participants to calibrate their models to the same set of
observational data. These data included the galaxy stellar mass function at
$z=0$ and $z=2$, the star formation rate function at $z=0.15$, the black-hole
bulge-mass relation at $z=0$, and the cold gas mass fraction at $z=0$.
Participants were free to weight these five calibrations as they saw fit, and
were asked to generate their ``best-fit'' model that took all of them into
account, i.e. calibration set `-c02' in Paper I.

We will build on previous work by investigating the evolution of the SMF for
the eight SAMs and one HOD model that were used in Paper I. These models are all
calibrated to the same observational data and are all run on the same background
dark matter only simulation, which means that we can discount the differences
due to the underlying cosmological framework when considering the differences
between the models. Our aim is then to see if the current physical prescriptions
used in any of the galaxy formation models can produce a realistic population of
galaxies at both low and high redshift. We will investigate the evolution of the
SMF in the redshift range $0.5 < z < 3.0$ for all nine galaxy formation models
and determine if models still struggle to simultaneously match observations both
at low and high redshift.

The rest of this paper is structured as follows: in Section \ref{sect:data} we
will briefly explain the underlying dark matter simulation and the parameters
used. In Section \ref{sect:results} we will present the results for the
evolution of the SMF and the passive fraction. We will then show how the
specific star formation rate of star-forming galaxies in the models evolves. We
will also examine the average halo mass as a function of stellar mass and the
stellar mass - halo mass relation for all the models. In Section
\ref{sect:discussion} we will present our discussion and in Section
\ref{sect:conclusions} we will present our conclusions.

\section{Simulation Data} \label{sect:data}
The eight SAMs we will be using are \DLB\ \citep{ref:DeLucia07}, \galform\
\citep{ref:GonzalezPerez14}, \galics\ \citep[][although the exact version used
for this comparison is the one described in the appendix of
\citet{ref:Knebe15}]{ref:Cattaneo17}, \LGal\ \citep{ref:Henriques13}, \morgana\
\citep{ref:Monaco07}, \sag\ \citep{ref:Cora18a}, \sage\ \citep{ref:Croton16} and
\ysam\ \citep{ref:Lee13}. The single HOD model is \mice\
\citep{ref:Carretero15}. A brief description of the physical prescriptions used
in each model is given in the Appendix of \citet{ref:Knebe15}. Any changes to
any of the models since then are included in Appendix \ref{app:models}. 

The models have all been run on the same underlying dark matter simulation which
may be different to the one used in the above reference papers. This can lead to
changes in the predictions of each model, as can varying the initial mass
function, yield, stellar population synthesis model and calibration data set
used.

A description of how the models were calibrated to the same observational data
is given in Paper I. We also note that the stellar masses from the models
have been convolved with a $0.08(1+z)$ dex scatter to account for the
observational errors when measuring stellar mass. This value comes from
\citet{ref:Conroy09}, who estimate an error of $\sim 0.2$ dex at $z=2$ when
fixing the stellar population synthesis model.

The underlying cosmological dark-matter-only simulation was run using the
\textsc{Gadget-3} $N$-body code \citep{ref:Springel05} with parameters given by
the Planck cosmology \citep[][$\Omega_{\rm m}=0.307$, $\Omega_\Lambda=0.693$,
$\Omega_{\rm b}=0.048$, $\sigma_8=0.829$, $h=0.677$, $n_s=0.96$]{ref:Planck13}.
We use $512^3$ particles of mass $1.24\times10^9$\hMsun\ in a box of comoving
width $125$\hMpc. The halo catalogues were extracted from 125 snapshots and
identified using \rockstar\ \citep{ref:Behroozi13a}. The halo merger trees were
then generated using the \textsc{ConsistentTrees} code \citep{ref:Behroozi13b}.

\section{Results} \label{sect:results}

\begin{figure*}
    \includegraphics[width=\textwidth]{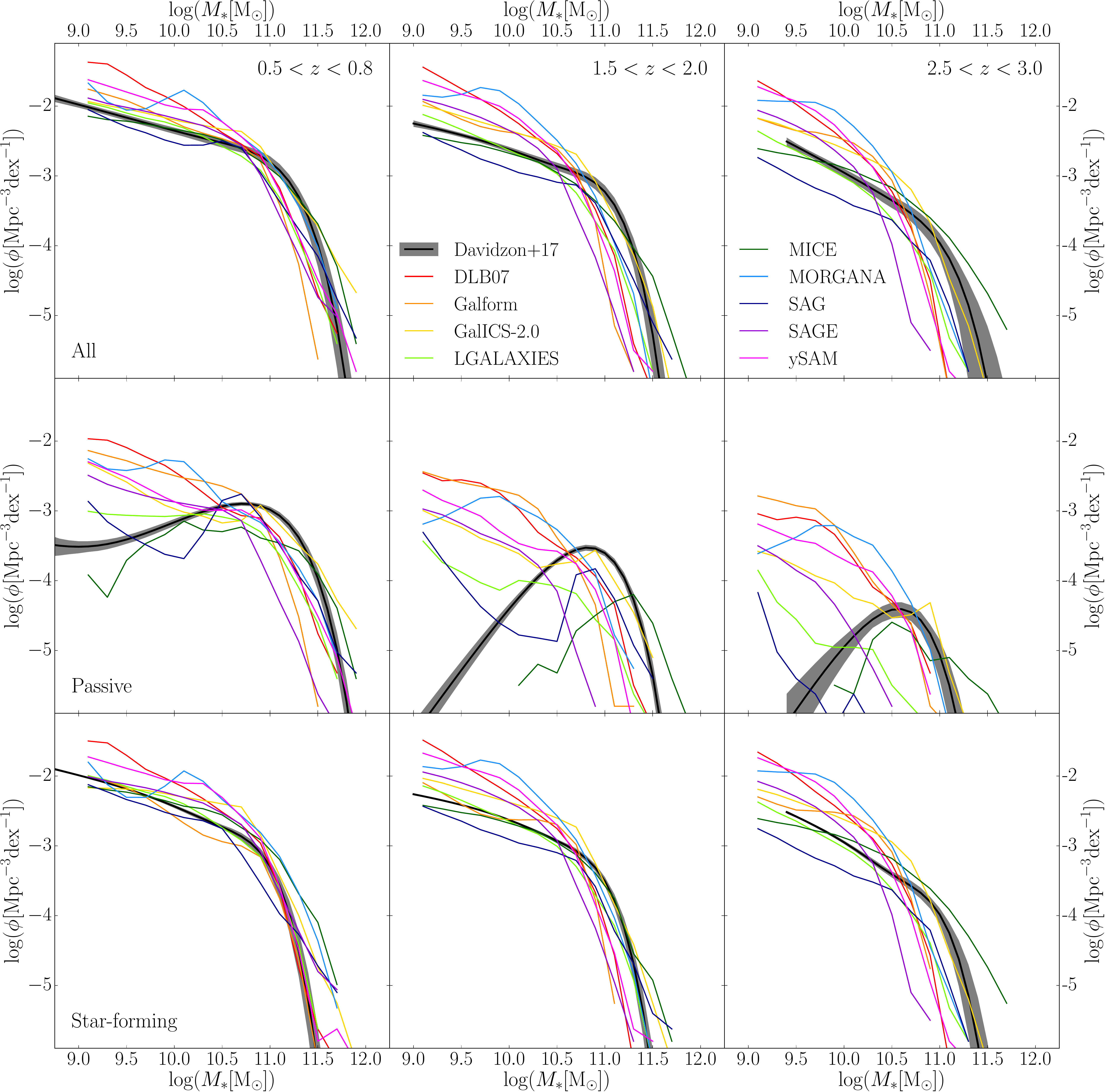}
    \caption[width=\textwidth]{The evolution of the stellar mass function for
all the models over the range $0.5 < z < 3.0$. The stellar mass function for the
whole, passive and star-forming samples are shown in the top, middle and bottom
panels, respectively, as coloured lines. The black line is the observational
best-fit mass functions from \citet{ref:Davidzon17}, with the dark grey shaded
region showing the $1 \sigma$ errors. For the models, each redshift bin contains
one snapshot, at redshifts $z=0.8, 2.0$ and $3.0$ respectively. We can see that
the models match well at low redshift (by construction), but deviate further
from the observations at high redshift. The number density of the lowest mass
objects is nearly constant in the models but changes by more than $0.5$ dex in
the observations. Most of the low-mass galaxies that are not present in the
observations at high redshift seem to be star-forming.}
    \label{fig:mass_functions}
\end{figure*}

\begin{figure*}
    \includegraphics[width=\textwidth]{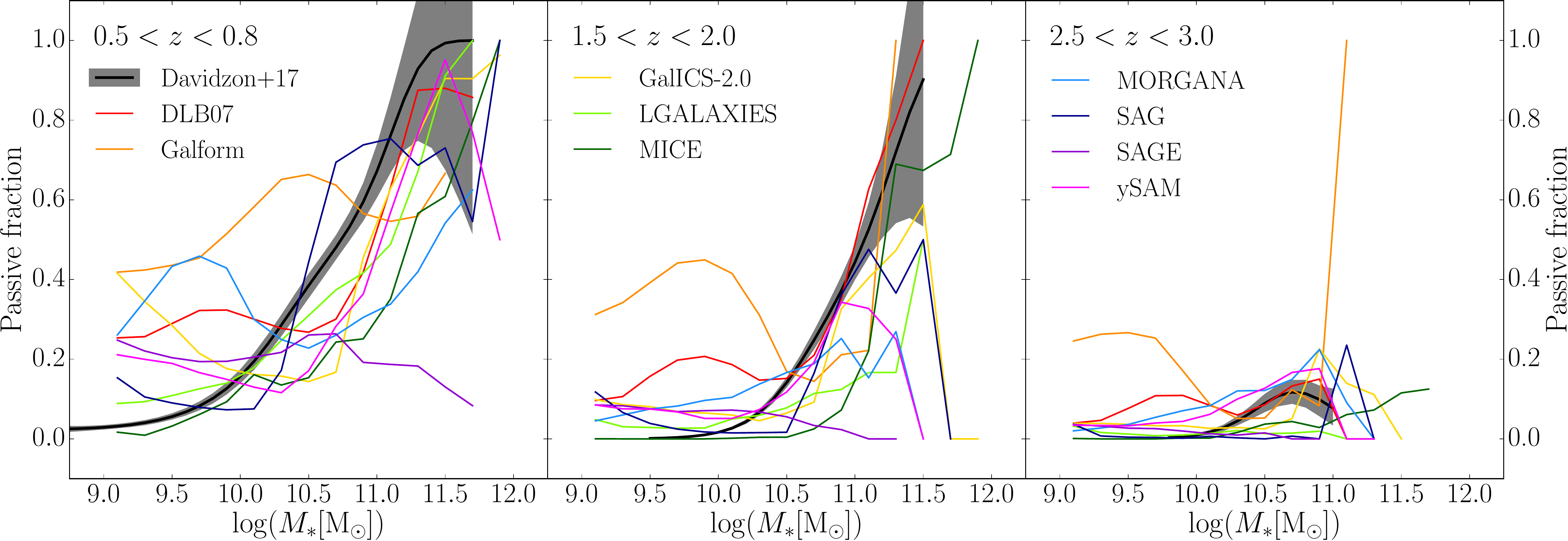}
    \caption[width=\textwidth]{The evolution of the passive fraction over the
range $0.5 < z < 3.0$. The coloured lines, black solid lines and grey shaded
regions are the same as in \Fig{fig:mass_functions}, as are the snapshots used
in each redshift bin for the models. For a few models the passive fraction is
too high at low masses, particularly at low redshift. The models match well at
high masses at low redshift, but generally underpredict the passive fraction for
high-mass galaxies at high redshift.}
    \label{fig:passive_fractions}
\end{figure*}

\begin{figure}  
    \includegraphics[width=\columnwidth]{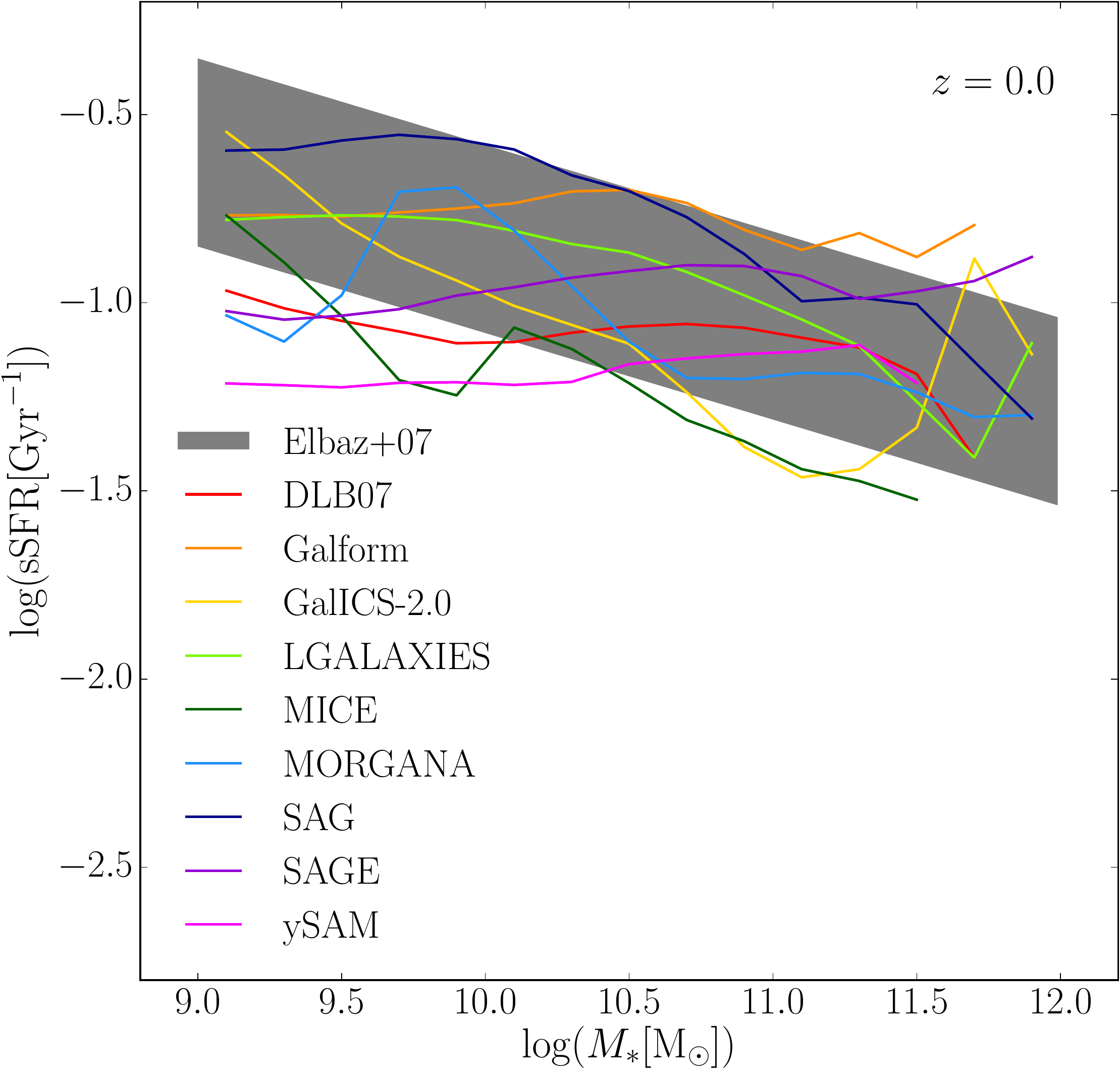}
    \caption[width=\columnwidth]{The relationship between mass and sSFR at
$z=0.0$ for star-forming galaxies in the nine models. The model data is taken
from one snapshot at $z=0.0$. The grey shaded region is taken from
\citet{ref:Elbaz07} and shows the observational best-fit to this relation. The
sSFR of star-forming galaxies in the models matches observations well but there
is less of a trend with mass in some models.}
    \label{fig:mass_ssfr_z_0}
\end{figure}

\begin{figure}  
    \includegraphics[width=\columnwidth]{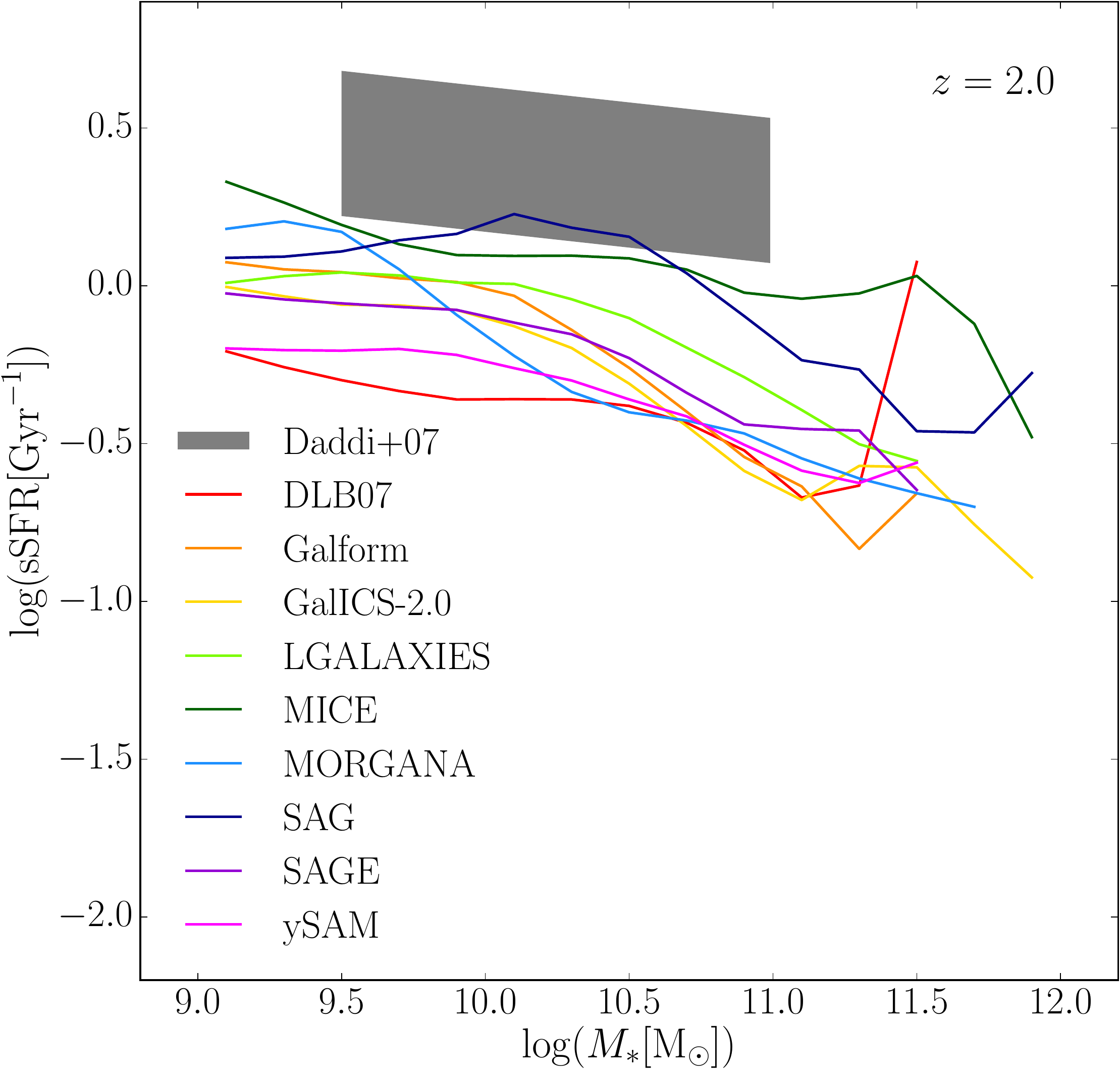}
    \caption[width=\columnwidth]{As for \Fig{fig:mass_ssfr_z_0}, but for $z=2.0$
and observations from \citet{ref:Daddi07}. The model data is taken from one
snapshot at $z=2.0$. Here all of the models lie almost completely below the
observational best-fit range.}
    \label{fig:mass_ssfr_z_2}
\end{figure}

\begin{figure*}  
    \includegraphics[width=\textwidth]{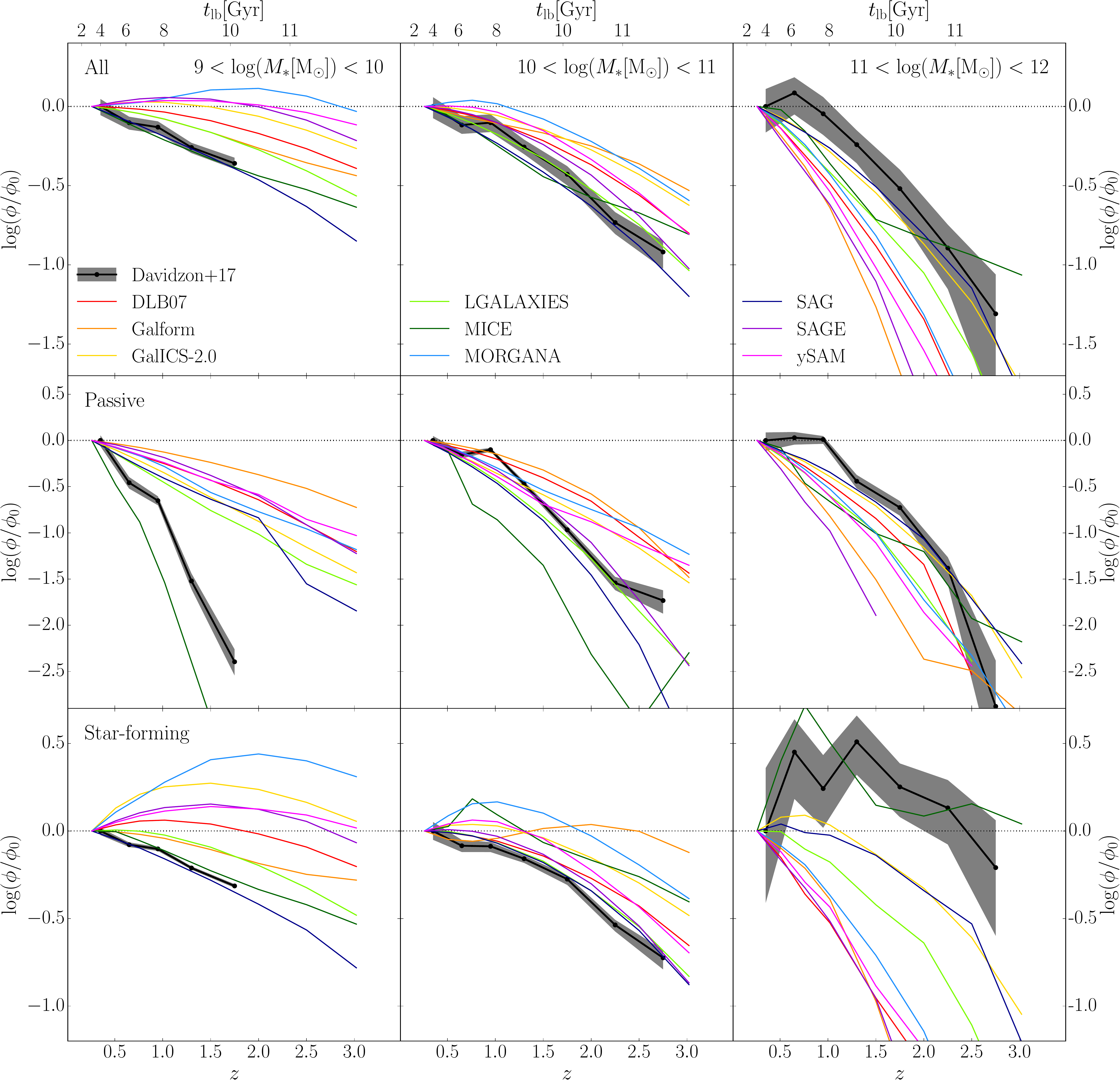}
    \caption[width=\columnwidth]{The evolution of the number density $\phi$, in
bins of stellar mass. This is normalised by the number density at $0.2 < z <
0.5$, which we call $\phi_0$. We show three mass bins as indicated (left to
right panels) for all galaxies (top panels), passive galaxies (middle panels)
and star-forming galaxies (bottom panels). The dark grey shaded regions and
black lines with circular points show data from \citet{ref:Davidzon17}. The
coloured points and lines are for the nine models. In the lowest mass bin, most
of the models assemble the galaxies before the observations. The models match
the observations well at intermediate masses, but the observational number
density increases before many of the models at high mass.}
    \label{fig:mass_function_evolution}
\end{figure*}

\begin{figure*}
    \includegraphics[width=\textwidth]{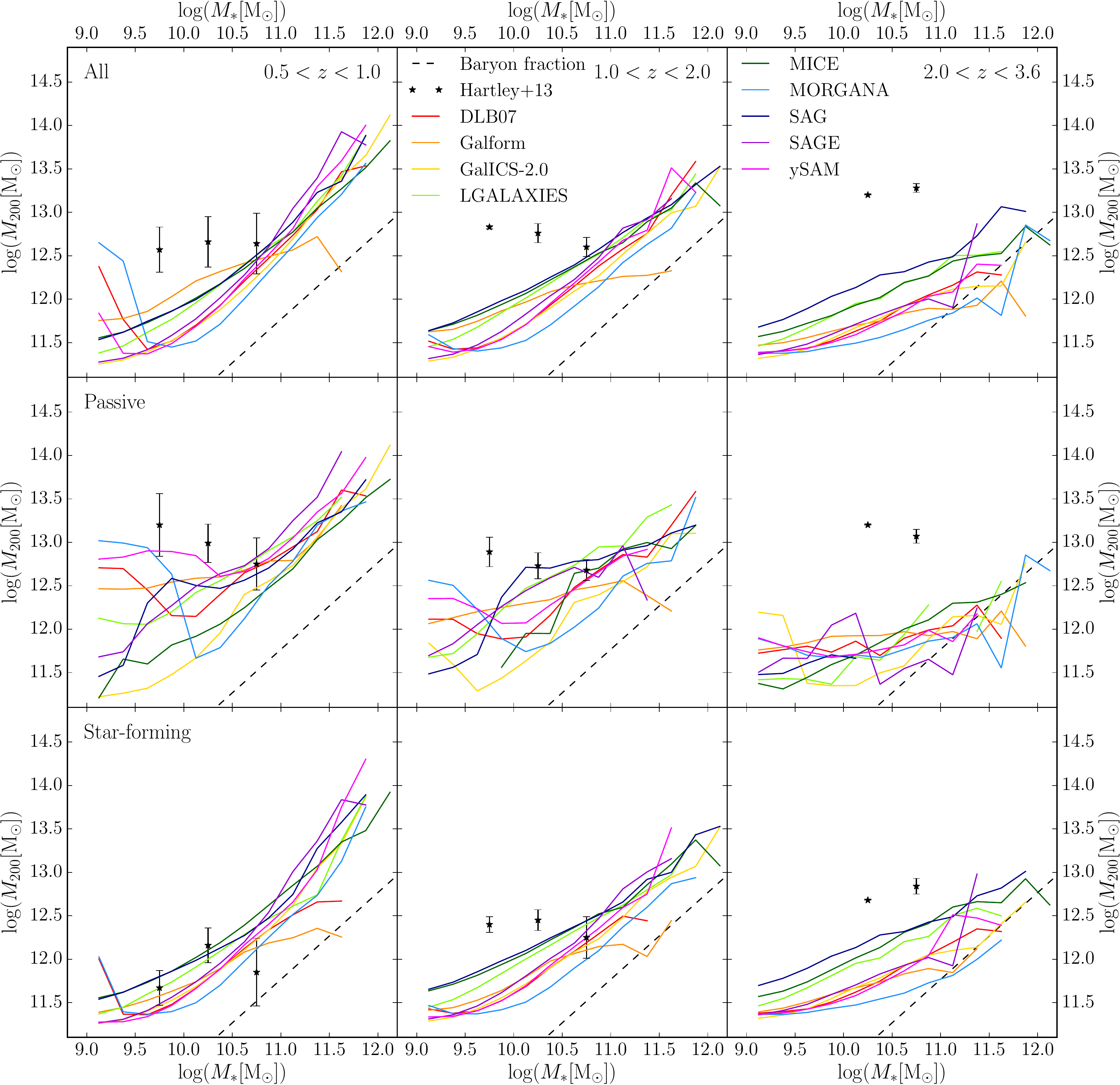}
    \caption[width=\textwidth]{The average halo mass in the models compared to
measurements from observations between $z=0.5$ and $z=3.6$. Observational
measurements of the average halo mass from \citet{ref:Hartley13} are derived
from clustering and are shown as stars. The values for each model are shown as
coloured lines. For the models, the mass of the main host halo was used rather
than the subhalo, to better compare with observational halo mass measurements
from clustering. The top panels cover the full galaxy sample, the middle panels
are for passive galaxies and the bottom panels are for star-forming galaxies.
The black dashed line shows the universal baryon fraction. For the models, we
use snapshots at $z=1.0, 2.0$ and $3.5$ for each redshift bin respectively. For
the passive sample, the halo masses from observations are approximately
constant, but decrease by up to a factor of 10 in the models with increasing
redshift. For the star-forming sample, the observations show halo mass
increasing with increasing redshift, whereas in the models there is no real
trend with redshift.}
    \label{fig:avg_halo_mass}
\end{figure*}

\begin{figure*}
    \includegraphics[width=\textwidth]{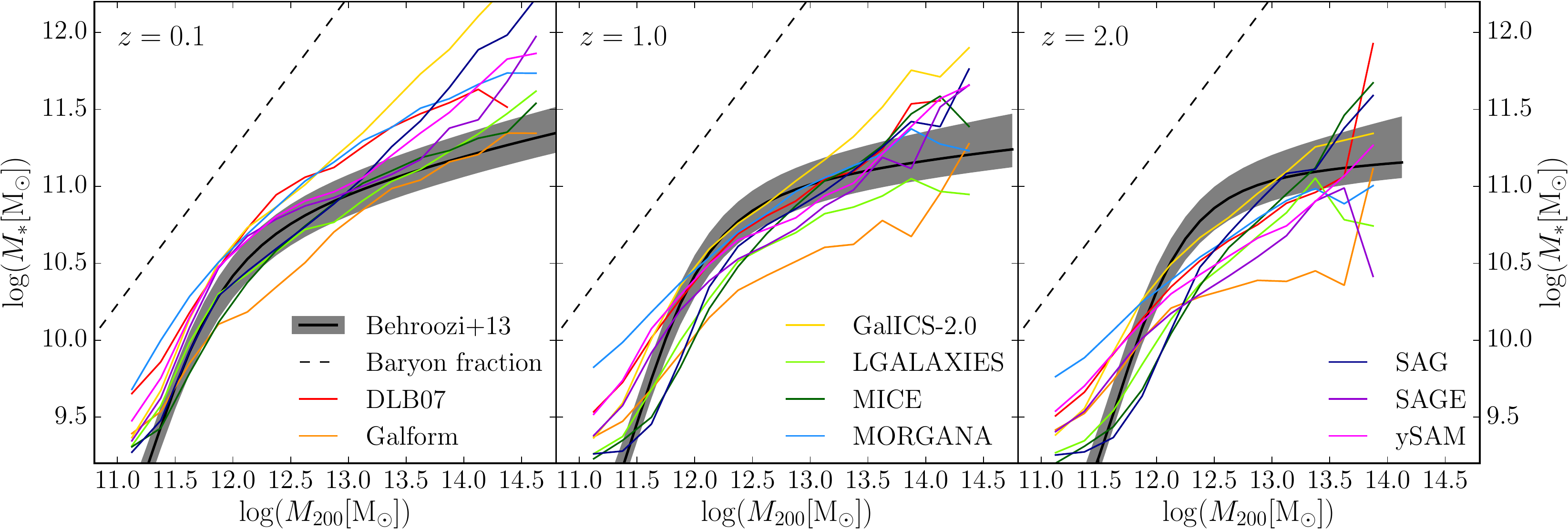}
    \caption[width=\textwidth]{Comparison of the average stellar mass for each
halo mass bin in the models to the abundance matching model of
\citet{ref:Behroozi13c}, considering only central galaxies. The results from the
models are shown as coloured lines and the stellar mass - halo mass relation
from \citet{ref:Behroozi13c} is shown as a dark grey shaded region and black
line. The black dashed line shows the universal baryon fraction. The panels are
for each redshift, increasing from left to right, using snapshots at $z=0.0,
1.0$ and $2.0$ respectively. Looking at the data from \citet{ref:Behroozi13c},
we can see that the average stellar mass stays fairly constant with redshift,
but increases in the models towards low redshift, particularly at high masses.}
    \label{fig:smhm_relation}
\end{figure*}

\subsection{Evolution of the Galaxy Stellar Mass Function}
\label{subsect:results_smf}

We start by examining the evolution of the stellar mass function in
\Fig{fig:mass_functions}, shown for the whole sample in the top row. The
coloured lines are the stellar mass functions for each of the models, computed
for each redshift bin using single snapshots at $z = 0.8, 2.0$ and $3.0$ for
each redshift bin respectively. We note that the precise choice of snapshot does
not affect our conclusions. The observations from \citet{ref:Davidzon17} are
based on the UltraVISTA near-infrared survey of the COSMOS field and are shown
as a black line and dark shaded region. When finding the best-fit Schechter
parameters to their stellar mass functions, they take into account the errors in
measuring stellar mass, known as Eddington bias. As they have applied this
correction, when plotting the stellar mass function we do not apply the
$0.08(1+z)$ dex scatter to the stellar mass values. In Appendix
\ref{app:convolved_smf} we have included a version of \Fig{fig:mass_functions}
where the model stellar masses do have this scatter applied, to show the
differences to the SMF.

We also note that we compare to different observational data than the combined
dataset used to calibrate the models. The data from \citet{ref:Davidzon17} is
more recent than the calibration dataset and also allows us to split our sample
into passive and star-forming galaxies. When comparing \citet{ref:Davidzon17} to
the stellar mass function calibration data at $z=0$ and $z=2$, the two largely
agree, although the former has smaller error bars. This is encouraging as it
shows good agreement between different observations.

Inspecting the top panels, what is clear is that the observational number
counts are evolving, with the high-mass end largely in place by $z=3$, while the
low-mass end rises at late times. Whilst the models match the observations
well at low redshift, the strong evolution at the low-mass end is not seen for
most of the galaxy formation models. The exceptions to this are \mice, \LGal\
and \sag, which all show an increasing number density of low-mass galaxies
towards low redshift. As \mice\ is an HOD model it has been designed to match
the evolution of the SMF. \LGal\ and \sag\ likely do a better job of matching
the SMF at high redshift due to the physics involved in the treatment of gas.
Both follow the prescription suggested in \citet{ref:Henriques13} of scaling the
reincorporation timescale of ejected gas with the inverse of the halo mass.
This means the process of gas being reincorpoated back into the halo takes
longer for low-mass haloes, shifting the growth of galaxies in these haloes from
early to late times. \sag\ also scales the reheated and ejected mass with
redshift to make supernova feedback more efficient at high redshift.

At the high-mass end, the models underestimate the number density compared to
observations, with \mice\ and \galics\ as the exceptions. One alternative reason
for this tension at the high-mass end may be due to \citet{ref:Davidzon17}
underestimating their uncertainties when accounting for Eddington bias, as it is
very difficult to accurately measure all of the sources of error. Due to the
steep slope of the SMF at high masses this would have a greater impact at the
high-mass end of the SMF. The impact of Eddington bias on the SMF are discussed
further in Appendix \ref{app:convolved_smf}.

\subsection{Star-forming and Passive Galaxy Stellar Mass Functions}
\label{subsect:results_smf_split}

We explore the mass growth further in the bottom two rows of
\Fig{fig:mass_functions}, splitting the population into passive (middle row) and
star-forming (bottom row) galaxies. We separate passive and star-forming
galaxies using a redshift-dependent specific star formation rate (sSFR) cut of
$\mathrm{sSFR}(z) = 1 / (3t_\mathrm{H}(z))$ where $t_\mathrm{H}(z)$ is the
Hubble time at that redshift. We test the robustness of this cut by examining
the change in our results when using slightly different cuts of
$\mathrm{sSFR}(z) = 1 / (2t_\mathrm{H}(z))$ and $\mathrm{sSFR}(z) = 1 /
(4t_\mathrm{H}(z))$. We find that the shape of the stellar mass function changes
very little and makes no difference to any of the conclusions that we draw. In
the observations the passive and star-forming galaxies are seperated using the
(NUV - r) vs (r - J) colour-colour diagram as described in \citet{ref:Ilbert13},
which is best suited to differentiate fully quiescent galaxies from those with
residual star formation. In practice, the exact location of the split makes
little difference to the low-mass end of the star-forming SMF and the high-mass
end of the passive SMF, as these galaxies will have very blue and red colours
respectively.

Splitting the galaxy population in this way reveals that the main source of the
difference between the observations and the models comes from the star-forming
population: low-mass star-forming galaxies appear to be far too common at high
redshift in the models and the star-forming SMF evolves little from $z=3$ to
$z=0.5$. The exceptions to this are \mice\ and \LGal, which appear consistent
with the observations at low masses up to $z=3$. For the passive galaxies, the
number density at low masses does evolve with redshift in the models, as seen in
the observations. However, most of the SAMs show rising number density towards
lower masses, in contrast with the observations which appear to show a turnover
or flattening of the passive SMF towards lower masses. In order to solve these
problems, models need to find a physically motivated way to reduce the star
formation rates of low-mass galaxies at high redshift. The same galaxies at
later times would then have lower stellar masses and star formation rates. This
would then act to redistribute the passive SMF in the models to better match the
observations.

\subsection{Evolution of the Passive Fraction}
\label{subsect:results_passive}

Another way of looking at this result is to examine the passive fraction, which
is shown in \Fig{fig:passive_fractions}. Again, the shaded regions indicate the
observations taken from the same source as used for \Fig{fig:mass_functions}.
The passive fraction indicates the ratio of passive to star-forming galaxies. At
low masses, some of the models, such as \DLB, \galform\ and \morgana, tend to
overestimate the passive fraction compared to observations. This has been seen
previously and appears to be linked to how environmental processes are taken
into account in the models \citep{ref:Lagos14, ref:GonzalezPerez18}. At low
redshift the number of star-forming galaxies matches observations well, so this
difference is due to the lack of a turnover or flattening of the passive SMF. At
higher redshifts, the overproduction of low-mass star-forming galaxies would act
to decrease the passive fractions. However, this is still too high in some
models, again due to the rising number density towards low masses in the passive
SMF. 

At low redshift, the models tend to match the observations well at high masses,
but one model, \sage, underpredicts the passive fraction. This is mainly due to
an underprediction for the number of high-mass passive galaxies. As shown by
\citet{ref:Stevens17}, detailing the structural evolution of galaxy discs with
the {\sc Dark Sage} variant of the model \citep{ref:Stevens16} leads to more
sensible passive fractions. In the redshift range $1.5 < z < 2.0$ the models
tend to underpredict the fraction of high-mass galaxies, mainly due to the lack
of high-mass passive galaxies above $z \sim 1$. The model which best matches the
observed passive fraction for high-mass galaxies is \DLB, which slightly
underpredicts the number density of both high-mass passive and star-forming
galaxies in this redshift range.

\subsection{Relationship between Mass and Specific Star Formation Rate}
\label{subsect:results_ssfr}

In order to better compare star formation in the observations and the models, we
also look at the specific star formation rates of the subset of star-forming
galaxies. \Fig{fig:mass_ssfr_z_0} shows the average sSFR as a function of mass
at $z = 0.0$ for each of the models as a solid coloured line. The grey shaded
region is taken from \citet{ref:Elbaz07}, who used SDSS data to find a fit to
the correlation between SFR and mass at $z = 0$. Their sample is made up of
19590 galaxies with redshifts $z=0.04-0.1$ and is complete to $\mathrm{M_B}
\leq -20$. \citet{ref:Brinchmann04} used $\mathrm{H}_\alpha$ emission to derive
the SFR of these galaxies and the stellar masses were derived by
\citet{ref:Kauffmann03}, who fit using a library of star formation histories to
find the most likely stellar mass.

Most of the models match the observations well here, with \LGal\ and \sag\ lying
in the observational region at all masses. Some models appear to evolve less
with mass than the observations suggest, with some showing almost no trend,
whereas the sSFR implied by the observations decreases by over 0.5 dex between
$10^9 \mathrm{M_\odot}$ and $10^{12} \mathrm{M_\odot}$. This means that some of
the models, such as \galform, match at low masses but not high masses, and
others such as \DLB\ and \ysam\ match at high masses but not low masses. This
was also discussed in \citet{ref:Guo16}, who used data from two SAMs, \galform\
and \LGal, and one hydrodynamical simulation, \eagle. They found that the median
sSFR remained almost constant with mass, in contrast with observations.

The relationship between sSFR and stellar mass at $z = 2.0$ is then shown in
\Fig{fig:mass_ssfr_z_2}. Here the observations are taken from
\citet{ref:Daddi07}, who use galaxies in the GOODS-S field to find the
correlation between SFR and mass at $z=2$. They are complete to $\mathrm{K}<22$
and use only $24 \mu \mathrm{m}$ selected galaxies in order to exclude passive
galaxies. The SFRs were estimated using the UV and the stellar masses were
derived by \citet{ref:Fontana04} using SED fitting.

This comparison highlights large differences between the observations and models
at this redshift, with the models almost completely outside the observational
range. The sSFR of star-forming galaxies in the models is on average around 0.5
dex lower than measured in the observations. The models therefore predict a
slower evolution of the sSFR with redshift than observations. This has been
previously seen by \citet{ref:Mitchell14}, who find that when they scale the
reincorporation time of gas with redshift they are able to better match the
evolution of the stellar mass function, but still underestimate the sSFR of
high-mass galaxies at $z \sim 2$. \citet{ref:Hirschmann16} also found that their
ejective models predicted lower than observed sSFRs at high redshift, even when
they could reproduce the growth of the stellar mass function.

Reducing the star formation rates of galaxies above $z \sim 2$, as suggested in
Section \ref{subsect:results_smf_split}, may help to solve this problem. If
galaxies have a lower star formation rate at higher redshift, their resulting
mass at lower redshift will be lower. A galaxy with the same star-formation
rate at $z=2$ will then have a higher sSFR as it will have a lower stellar mass.

\subsection{Growth of the Galaxy Stellar Mass Function}
\label{subsect:results_smf_growth}

In \Fig{fig:mass_function_evolution} we examine the growth of the stellar mass
function as a function of mass and redshift. This is found by taking the value
of the number density $\mathrm{\phi}$ at fixed stellar mass for a certain
redshift bin and normalising it by the value of $\mathrm{\phi}$ in the lowest
redshift bin $0.2 < z < 0.5$, which we call $\mathrm{\phi_0}$. This allows for
easier comparison between the models and observations and will highlight when
the number density of different populations increases. The dark grey region and
black line with circular points shows data from \citet{ref:Davidzon17}. The
coloured lines then show the number density evolution for the nine models. The
black dotted line shows where the number density is equal to the number density
in the lowest redshift bin.

Looking at the passive galaxies, we can see that the models struggle to match
the observed growth of the mass function at low masses, as the number density of
low-mass galaxies increases in the models at higher redshift than the
observations. The only exception is \mice, which has very few galaxies with mass
below $10^{10}\mathrm{M_\odot}$ above $z \sim 1$. At intermediate masses the
models match the observations well, but at high masses the growth of the mass
function occurs in observations before many of the models.

For the star-forming galaxies, at low masses there is a similar problem with
several of the models; the mass function grows too much at high redshift. Just
under half of the models have more low-mass star-forming objects in the highest
redshift bin than the lowest redshift one. However, several of the models are
more in line with the growth of the observed mass function, namely \sag\ and
\mice. At intermediate masses, the number density of star-forming galaxies
increases at higher redshift in the models than in the observations. The model
that is most discrepant, \morgana, has more intermediate mass star-forming
galaxies between $1.0 < z < 1.5$ than in the lowest redshift bin. For high-mass
galaxies, the number density evolves little since high redshift in the
observations. \mice\ reproduces this trend well but in other models the number
density increases at lower redshift. This may be in part due to the fact that
there are low numbers of the highest mass galaxies which will naturally
introduce more scatter in the proportional change in number density.

We can also see interesting differences between models when comparing the
stellar mass function to the growth of the stellar mass function. Looking
at the lower panel of \Fig{fig:mass_functions}, we can see that \DLB\
overpredicts the number of low-mass star-forming galaxies at both low and 
high redshift. \sage\ agrees well with observations at low redshift but
overproduces low-mass star-forming galaxies at high redshift. However, looking
at the lower left panel of \Fig{fig:mass_function_evolution}, we can see that
\DLB\ matches observations of the growth of the mass function better than \sage.
These models therefore have slightly different problems; \DLB\ has too many
low-mass galaxies at all redshifts, but the number density increases at the
correct rate. Conversely, \sage\ has the correct number at low redshift, but the
number density increases too early.

\subsection{Average Halo Mass}
\label{subsect:results_avg_halo}

In this section we study the average halo mass the galaxies reside within, shown
in \Fig{fig:avg_halo_mass}. For the models we use single snapshots at $z =
1.0, 2.0$ and $3.5$ for each redshift bin respectively. The dashed black line
indicates the universal baryon fraction, i.e. where all the baryonic material
within the halo has been converted into stars. Each of the coloured lines
indicates the average halo mass values for a different model, while the black
points with errorbars are average halo mass values taken from
\citet{ref:Hartley13}, who use the UDS DR8 data to estimate the halo masses from
measurements of galaxy clustering \citep[e.g.][and references
therein]{ref:Mo02}. 

For the models, here we use the mass of the main host halo for each galaxy
rather than the mass of its subhalo. Host haloes do not reside within another
halo, whereas subhaloes are contained within a host halo. Although using the
host halo is not necessarily the usual choice when analysing simulation data, it
allows us to compare to observational measurements of halo mass from galaxy
clustering, which effectively measure the mass of the main host haloes
\citep{ref:Mo02}. For this reason we also include both centrals and satellites,
in order to best mimic the observational measurements. Assuming galaxy
clustering measurements can correctly recover the host halo mass, we can then
directly compare the observations and models.

Splitting the sample into passive and star-forming galaxies in
\Fig{fig:avg_halo_mass} we see that there are marked differences between the
observations and the models. For passive galaxies, the average halo mass in
observations stays constant over redshift in the observations, but rises towards
low redshift in the models. For the star-forming population, while the
observations indicate a general downsizing trend in halo mass of about an order
of magnitude between high and low redshift, all the models show virtually no
change. It is clear that the models start significantly below the observations
at $2.0 < z < 3.5$ and only agree with the observations by $0.5 < z <
1.0$. Both passive and star-forming low-mass galaxies are therefore in lower
mass haloes on average in the models than in the observations at high redshift.

One thing that can affect the average halo mass values in the models is the halo
mass definition used, as this can lead to differences of up to 20 percent
\citep{ref:Jiang14}. Although this may account for some of the scatter between
the models, the differences between the observations and models cannot be
explained by this alone. Another factor that could affect the observational
measurments of halo mass from clustering is `halo assembly bias', which refers
to the fact that halo clustering can depend on other properties besides halo
mass. For example, \citet{ref:Gao05} found that at fixed halo mass, haloes that
assembled earlier are more clustered than those that assembled later. Therefore,
galaxies in older haloes will be more strongly clustered than they should be for
their halo mass, which means that their halo masses will be measured as higher
than they actually are. This could alleviate some of the discrepancy between the
observations and models. For example, if the passive galaxies observed at
low redshift are associated to older haloes, then their halo masses could have
been overestimated.

\subsection{Stellar Mass - Halo Mass Relation}
\label{subsect:results_smhm}

In \Fig{fig:smhm_relation} we display measurements of the average stellar mass
of central galaxies in bins of halo mass, comparing the models with the
abundance matching model of \citet{ref:Behroozi13c}. The dashed black line
indicates the universal baryon fraction and the dark grey region and black solid
line show the fit to the stellar mass - halo mass (SMHM) relation from
\citet{ref:Behroozi13c}. The coloured lines show the average stellar mass values
for each different model.

At low redshift, the results from the models and the SMHM relation agree well at
low and intermediate halo masses. However, above halo masses of $\sim
10^{13.5}\mathrm{M_\odot}$ the average stellar mass of centrals in the models is
higher than suggested by the SMHM relation. This means that at low redshift,
star formation in high-mass haloes is more efficient in the models. The
exceptions to this are \LGal\ and \mice, which agree with the SMHM relation at
nearly all halo masses. For most of the models, the slope of the relation at
high halo masses does flatten, but not to the extent seen from the SMHM
relation.

As we move to higher redshift the SMHM relation changes little. The peak of the
relation moves to slightly higher halo masses and the average stellar mass for
low-mass haloes decreases by $\sim 0.4$ dex at $10^{11.5}\mathrm{M_\odot}$. In
the models the average stellar mass for low-mass haloes decreases slightly with
increasing redshift, but is above the SMHM relation by $z=1.0$ for most models.
This discrepancy can be partially explained by the cut in stellar mass applied
at $M_* = 10^{9}\mathrm{M_\odot} h^{-1}$, which may have skewed the distribution
towards higher stellar masses. This might be enough to explain the difference
for models such as \LGal\ or \galform, but the discrepancy is too large for
\morgana, \DLB\ and \ysam. In these models, the average stellar mass for
low-mass haloes at high redshift is too high. This means that star formation in
these objects is very efficient, leading to an increase in the number of
low-mass galaxies at $z \sim 2$. This is likely due to the way that the physics
involved in the gas cycle is implemented in these models.

For intermediate- and high-mass haloes, the average stellar mass generally
decreases with increasing redshift in the models and the slope of the relation
decreases. This suggests that star formation was less efficient in the models at
high redshift. At $z=0.1$ the models overpredict the stellar mass in high-mass
haloes, but slightly underpredict it by $z=2.0$. For intermediate-mass haloes,
the average stellar mass is too low in the models at $z=2.0$ by up to 0.5 dex,
as is the case for \galform\ at $10^{12.5}\mathrm{M_\odot}$. The model that
changes the least with redshift is \mice; as this is an HOD model it naturally
matches the SMHM relation better than the SAMs.

\section{Discussion} \label{sect:discussion}
Comparing several galaxy formation models allows us to distinguish areas that
are challenging for the current generation of models and therefore provide
direction for the future development of the field as a whole. The main issue
highlighted in this paper is the fact that most of the models produce too many
low-mass, star-forming galaxies at early times. Observationally these appear
either to not exist or to be missed by the surveys. This is a difficult area
observationally with the answer to this question only becoming evident when the
stellar mass functions are reliably pushed to lower masses. At present they are
tantalisingly close to indicating a clear turnover in the space density of
passive galaxies at low-mass, which would significantly challenge many of the
models featured here.

In the absence of a new population of low-mass, star-forming galaxies being
observed at $z \sim 2$, many of the models would need improvements in order to
reproduce observations. They would need to produce far fewer low-mass
star-forming galaxies at essentially all but the latest times. Shifting star
formation from high-mass haloes at high redshift to low-mass haloes at
low redshift would also produce better agreement with observations of galaxy
clustering. Reducing the number of low-mass star-forming objects would also have
to be achieved without reducing the number of high-mass objects significantly.

Some of the models, such as \LGal\ and  \sag, do fit the low-mass end of both
the star-forming and passive stellar mass function at high redshift. This is
likely due to their implementation of the physics involved in the treatment of
gas, in particular the reincorporation timescales. \mice\ also matches
observations at high redshift, but as this is a HOD model it matches by
construction. However, there are still some observables that even these models
struggle to match, such as the relation between stellar mass and specific star
formation rate and the average halo mass that galaxies occupy. Whilst this could
be due to problems with the observational measurements of these quantities, this
could point towards areas where the models still need to improve.

\section{Conclusions} \label{sect:conclusions}
In this paper we have contrasted nine different galaxy formation models and
compared them to the latest high-redshift observations. In doing so we have
highlighted the areas in which the models find particular difficulty in matching
the observations. We can see from this project that some of the models still
have trouble simultaneously matching the stellar mass function at both low and
high redshift. The galaxies look roughly correct at $z=0$, but for many models
there are too many low-mass galaxies at $z \sim 2$, as has also been seen
previously \citep[e.g.][]{ref:Fontana06, ref:Fontanot09, ref:Weinmann12,
ref:Henriques12, ref:Guo16}.

To explore this further, we split galaxies into passive and star-forming
populations. We find that there are too many star-forming galaxies with
stellar masses below $10^{11}\mathrm{M_\odot}$ in many of the models at $z \sim
2$.

In summary, while some of the models are remarkably successful at reproducing
the evolution of the stellar mass function, there remain significant issues. In
particular:

\begin{itemize}
  \item Whilst most of the models are able to match the observed stellar mass
function at low redshift, they tend to overproduce the number density of
low-mass galaxies at high redshift.
  \item In most of the models the low-mass end of the star-forming stellar mass
function is already largely in place at high redshift $(z>1)$, in contrast to
observations. This is because the models appear to produce too many star-forming
galaxies below the knee of the stellar-mass function at early times.
  \item The passive stellar mass function from the models evolves with redshift
as in the observations, but does not have the same turnover or flattening in the
number density at the low-mass end.
  \item Whilst most of the models match the passive fraction well at high
masses, for some of the models the passive fraction is too high at low masses.
This is despite the overproduction of low-mass star-forming galaxies.
  \item Most of the models are able to reproduce the relationship between sSFR
and the mass of the star-forming galaxies at low redshift, but underpredict the
sSFR at high redshift.
  \item Observational measurements of halo mass, estimated from galaxy
clustering, indicate clear downsizing in the average halo mass occupied by
star-forming galaxies as a function of redshift. This is not clearly indicated
by any of the models; both star-forming and passive galaxies in the models
occupy haloes with lower masses than those inferred from observations at $z=2$.
  \item The average stellar mass is higher in low-mass haloes at high redshift
in the models compared to observations, meaning that star formation in low-mass
haloes is more efficient in the models than in the real Universe.
\end{itemize}

Achieving consistent results at both $z=0$ and $z=2$ with a population of
galaxies that evolves strongly with redshift is clearly difficult. The HOD
model, \mice, obtains good results but the galaxies present at $z=2$ are not
evolved directly into the $z=0$ population. Of the SAMs, the \LGal\ and \sag\
models best match the growth of the observed mass functions, but they share the
same trends as the other models for the specific star formation rate and average
halo mass within which the objects reside. Both of these models found that they
needed to modify the treatment of the gas cycle in order to match the evolution
of the low-mass end of the stellar mass function. This is very promising for the
galaxy formation modelling community, which has long struggled with this issue.

While it is clear that current galaxy formation models can reproduce a variety
of observational data, we have identified key areas of tension. Some models
still overpredict the number of low-mass galaxies at high redshift, but even
the models that can match the evolution of the galaxy stellar mass function
underpredict the specific star formation rates of galaxies at early times.
Future observational surveys at high redshift will help shed light on these
issues and identify further areas of improvement for the models.

\section*{Acknowledgements}
We thank the anonymous referee whose comments helped to greatly improve this
paper. We would also like to thank Rachel Somerville, Gabriella De Lucia and
Pierluigi Monaco for kindly providing useful discussion and comments.

We thank Carnegie Observatories for their support and hospitality during the
workshop `Cosmic CARNage' where all the calibration issues were discussed and
the roadmap laid out for the work presented here.

The authors would further like to express special thanks to the Instituto de
Fisica Teorica (IFT-UAM/CSIC in Madrid) for its hospitality and support, via the
Centro de Excelencia Severo Ochoa Program under Grant No. SEV-2012-0249, during
the three week workshop `nIFTy Cosmology' where this work developed. We further
acknowledge the financial support of the 2014 University of Western Australia
Research Collaboration Award for `Fast Approximate Synthetic Universes for the
SKA', the ARC Centre of Excellence for All Sky Astrophysics (CAASTRO) grant
number CE110001020, and the ARC Discovery Project DP140100198. We also recognise
support from the Universidad Autonoma de Madrid (UAM) for the workshop
infrastructure.

RA is funded by the {\it Science and Technology Funding Council} (STFC) through
a studentship.
AK is supported by the {\it Ministerio de Econom\'ia y Competitividad} and the
{\it Fondo Europeo de Desarrollo Regional} (MINECO/FEDER, UE) in Spain through
grant AYA2015-63810-P. He further thanks Denison Witmer for california brown and
blue.
VGP acknowledges support from a European Research Council Starting Grant
(DEGAS-259586). This work used the DiRAC Data Centric system at Durham
University, operated by the Institute for Computational Cosmology on behalf of
the STFC DiRAC HPC Facility (www.dirac.ac.uk). This equipment was funded by BIS
National E-infrastructure capital grant ST/K00042X/1, STFC capital grant
ST/H008519/1, and STFC DiRAC Operations grant ST/K003267/1 and Durham
University. DiRAC is part of the National E-Infrastructure. 
FJC acknowledges support from the Spanish Ministerio de Econom\'ia y
Competitividad project AYA2012-39620.
SAC acknowledges funding from {\it Consejo Nacional de Investigaciones
Cient\'{\i}ficas y T\'ecnicas} (CONICET, PIP-0387), {\it Agencia Nacional de
Promoci\'on Cient\'ifica y Tecnol\'ogica} (ANPCyT, PICT-2013-0317), and {\it
Universidad Nacional de La Plata} (G11-124), Argentina.
DJC acknowledges receipt of a QEII Fellowship from the Australian Government.
FF acknowledges financial contribution from the grants PRIN MIUR 2009 `The
Intergalactic Medium as a probe of the growth of cosmic structures' and PRIN
INAF 2010 `From the dawn of galaxy formation'.
The work of BH was supported by Advanced Grant 246797 GALFORMOD from the
European Research Council.
NDP was supported by BASAL PFB-06 CATA, and Fondecyt 1150300.  Part of the
calculations presented here were run using the Geryon cluster at the Center for
Astro-Engineering at U. Catolica, which received funding from QUIMAL 130008 and
Fondequip AIC-57.
CP acknowledges support of the Australian Research Council (ARC) through Future
Fellowship FT130100041 and Discovery Project DP140100198. WC and CP acknowledge
support of ARC DP130100117.
PAT acknowledges support from the Science and Technology Facilities Council
(grant number ST/L000652/1).
SKY acknowledges support from the Korean National Research Foundation
(NRF-2017R1A2A1A05001116). This study was performed under the umbrella of the
joint collaboration between Yonsei University Observatory and the Korean
Astronomy and Space Science Institute. The supercomputing time for the numerical
simulations was kindly provided by KISTI (KSC-2014-G2-003).

The authors contributed to this paper in the following ways: RA analysed the
data, created the plots and wrote the paper along with FRP and OA.  AK \& CP
formed part of the core team and along with FRP organised the nIFTy workshop
where this work was initiated. AB organised the follow-up workshop `Cosmic
CARNage' where all the discussions about the common calibration took place and
out of which this paper emerged. JO supplied the simulation, halo catalogue and
merger tree for the work presented here. WH supplied the halo mass
measurements from the UDS that were used in this work. The remaining authors
performed the SAM or HOD modelling using their codes, in particular FJC, AC, SC,
DC, FF, VGP, BH, JL, ARHS, CVM, and SKY actively ran their models. All authors
proof-read and commented on the paper.

This research has made use of NASA's Astrophysics Data System (ADS) and the
arXiv preprint server.



\bibliographystyle{mnras}
\bibliography{references} 



\appendix

\section{Galaxy Formation Models} \label{app:models}

A description of the physical prescriptions of each model is available in the
Appendix of \citet{ref:Knebe15}. Here we present a brief description of the
changes to any of the models since then:


\subsection{SAG}

The changes implemented in SAG are described in detail in \citet{ref:Cora18a}.
We summarize them here: 

\paragraph*{\it Cooling}
Both central and satellite galaxies experience gas cooling processes. Satellite
galaxies keep their hot gas haloes which are gradually removed by the action of
ram pressure stripping (RPS), modelled according to \citet{ref:McCarthy08}, and
tidal stripping (TS). When the mass of the hot gas halo becomes smaller than
$10$ percent of the total baryonic mass of the galaxy, it is assumed that it no
longer shields the cold gas disc from the action of RPS, which is modelled
following the criterion from \citet{ref:Gunn72}; see \citet{ref:Tecce10} for
more details. Values of ram pressure experienced by galaxies in haloes of
different mass as a function of halo-centric distance and redshift are obtained
from fitting formulae derived from the self-consistent information provided by
the hydrodynamical simulations analysed by \citet{ref:Tecce10}, as described in
Vega-Mart\'inez et al. (in prep.).

\paragraph*{\it Supernova feedback and winds}
The mass reheated by supernova feedback involves an explicit redshift dependence
and an additional modulation with virial velocity, according to a fit to results
from FIRE (Feedback in Realistic Environments) hydrodynamical simulations
\citep{ref:Muratov15}.

\paragraph*{\it Gas ejection and reincorporation}
The energy input by massive stars eject some of the hot gas out of the halo,
according to the energy conservation argument presented by \citet{ref:Guo11}.
The energy injected by massive stars is proportional to the mean kinetic energy
of supernova ejecta per unit mass of stars formed, and includes the same
explicit redshift dependence and the additional modulation with virial velocity
as the reheated mass. The ejected gas mass is re-incorporated back onto the
corresponding (sub)halo within a timescale that depends on the inverse of the
(sub)halo mass \citep{ref:Henriques13}.

\paragraph*{\it AGN feedback}
AGN are produced from the growth of central BHs. When this growth takes place
from cold gas accretion during gas cooling, it depends on the mass of the hot
gas atmosphere, following \citet{ref:Henriques15}.

\paragraph*{\it Orphans}
The positions and velocities of orphan galaxies are obtained from the
integration of the orbits of subhaloes that will not longer be identified.
The orbits are integrated numerically, considering the last known position,
velocity and virial mass of subhaloes as initial conditions, and taking into
account mass loss by TS and dynamical friction effects, following some aspects
of the works by \citet{ref:Gan10} and \citet{ref:Kimm11}. A merger event occurs
when the halo-centric distance becomes smaller that $10$ percent of the virial
radius of the host halo.


\subsection{SAGE}

The only change in \sage\ is to the radio mode AGN feedback. It is explained in
detail in \citet{ref:Croton16} and summarized here:

\paragraph*{\it AGN feedback}
The radio mode AGN feedback has been modified in \sage\ since
\citet{ref:Croton06}. There is now a heating radius, inside which gas is
prevented from cooling. This heating radius increases with subsequent heating
episodes and can not decrease.


\section{Stellar Mass Function Including Eddington Bias}
\label{app:convolved_smf}

In \Fig{fig:mass_functions} we have compared the stellar mass function from the
models to observational data from \citet{ref:Davidzon17}. We do not scatter the
stellar masses in the models with the $0.08(1+z)$ dex scatter used to mimic
observational uncertainties, as \citet{ref:Davidzon17} have accounted for this
when finding the best-fit Schechter parameters to their stellar mass function.

\begin{figure*}
    \includegraphics[width=\textwidth]{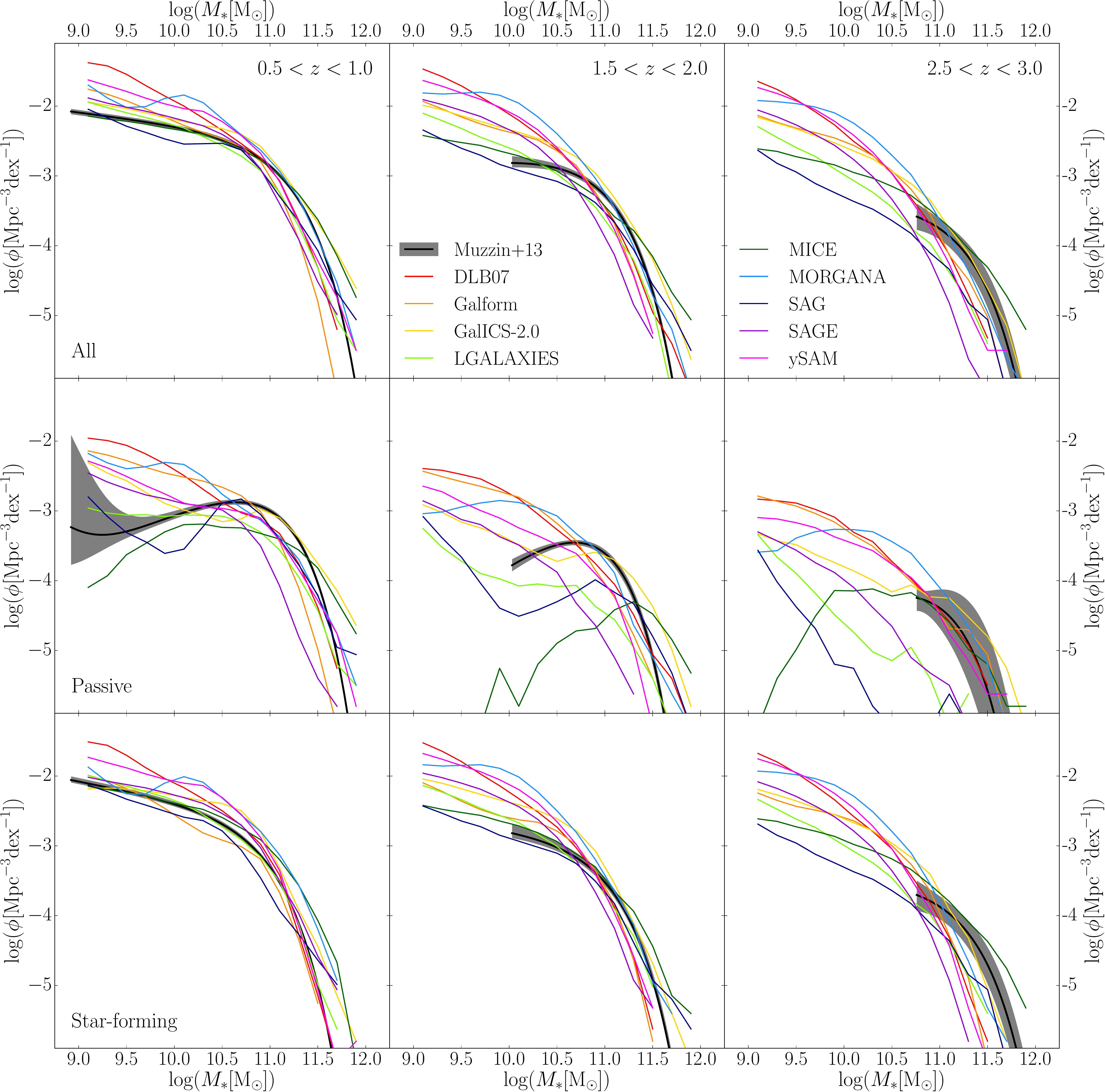}
    \caption[width=\textwidth]{Alternative version of \Fig{fig:mass_functions},
applying the $0.08(1+z)$ dex scatter to the stellar mass values in the models.
Here we compare to observational data from \citet{ref:Muzzin13}, who do not
take into account Eddington bias when finding the best-fit Schechter
parameters. Here the models match the observations better at high masses
and high redshift.}
    \label{fig:mass_functions_convolved}
\end{figure*}

Here we present an alternative version of \Fig{fig:mass_functions}, shown in
\Fig{fig:mass_functions_convolved}, where we do apply the scatter to the stellar
mass values in the models. We compare to observations from \citet{ref:Muzzin13},
who do not take these uncertainties into account when fitting to the stellar
mass function. Like \citet{ref:Davidzon17}, the observations from
\citet{ref:Muzzin13} are based on the UltraVISTA near-infrared survey of the
COSMOS field.

Comparing \Fig{fig:mass_functions_convolved} to \Fig{fig:mass_functions}, we can
see that the main difference to the SMF is at the high-mass end and that the
low-mass end is largely unaffected. Due to the redshift dependence of the
scatter we apply to the stellar masses, the differences are also larger at high
redshift. As an example, the value of $\phi$ increases by over 0.5 dex at
$10^{11}\mathrm{M_\odot}$ in the redshift bin $2.5 < z < 3.0$ in \LGal\ when
the scatter is applied.

In \Fig{fig:mass_functions}, it appears that most of the models underpredict the
number of high-mass galaxies at high redshift, with only \mice\ and \galics\
matching observations. However, in \Fig{fig:mass_functions_convolved} the
models and observations agree better at high redshift for several other models,
namely \galform\ and \morgana. The data from \citet{ref:Muzzin13} form part of
the combined dataset used to calibrate the models, so it is natural that the
models may match this data better.

\bsp	
\label{lastpage}
\end{document}